# Cloud based DevOps Framework for Identifying Risk Factors of Hospital Utilization

Monojit Banerjee
Senior Member
*IEEE*
Delaware, USA
monojitbanerjee@ieee.org

Akaash Vishal Hazarika
Department of Computer Science
North Carolina State University
Raleigh, USA
ahazari@alumni.ncsu.edu

Mahak Shah
Department of Computer Science
Columbia University
New York, USA
ms5914@caa.columbia.edu

*Abstract*—A scalable and reliable system is required to analyze the National Health and Nutrition Examination Survey (NHANES) data efficiently to understand hospital utilization risk factors. This study aims to investigate the integration of continuous integration and deployment (CI/CD) practices in data science workflows, specifically focusing on analyzing NHANES data to identify the prevalence of diabetes, obesity, and cardiovascular diseases. An end-to-end cloud-based DevOps framework is proposed for data analysis which examines risk factors associated with hospital utilization and evaluates key hospital utilization metrics. We have also highlighted the modular structure of the framework that can be generalized for any other domains beyond healthcare. In the framework, an online data update method is provided which can be extended further using both real and synthetic data. As such, the framework can be especially useful for sparse dataset domains such as environmental science, robotics, cybersecurity, and cultural heritage and arts.

*Keywords—Machine Learning, Cloud Computing, DevOps, Large Scale Systems, MLOps, Hospital Utilization*

## I. Introduction

This study investigates risk factors associated with hospital utilization and major diseases like diabetes, obesity, and cardiovascular disease. Utilizing data from the National Health and Nutrition Examination Survey (NHANES) spanning 1999-2016 [1], the analysis examines demographic, dietary, examination, laboratory, and questionnaire data about hospital utilization. As part of this study, an end-to-end cloud-based DevOps [2] framework is set up. This framework utilizes a data collection system, a source code repository system for continuous integration(CI), and finally a continuous deployment(CD) pipeline to deploy the analytical program in a cloud environment. Moreover, the data collection system is also integrated with the CI system which helps the framework adapt to any new data. For example- adding another year worth of data will automatically trigger required change and update the analytical result.

## II. Background

Hospital utilization is defined as the number of patients who are formally accepted as inpatients. How many times a patient received healthcare over the past year is an indication of that. The goal of this project is to identify which risk factors are associated with hospital utilization. Furthermore, prevalence [3] is defined as the proportion (%) of a particular population affected by the exposure or disease of interest. This project also aims to identify the prevalence of major diseases such as diabetes, obesity, and cardiovascular diseases. NHANES data was studied in many different contexts including prepandemic data files and presents prevalence estimates [4]. Also, NHANES is periodically updated and it often contains useful information that can influence public policy [5].

Continuous Integration (CI) and Continuous Delivery/Deployment (CD) are essential practices in modern software development. Some studies highlighted the challenges of deploying machine learning models in healthcare settings[6]. Also, machine learning models such as XGBoost can help with predicting 30-day hospital readmission [7]. Some studies [8], and [9] also compared specialized model performance with common ML algorithms in healthcare settings.

However, we have found that these studies are mostly ad-hoc analysis that uses an offline trained model or general discussion about how DevOps can be used in healthcare. In our research, we have aimed to solve these challenges by proposing a comprehensive system integrated with a DevOps pipeline that can use both online and offline data and continuously train and update the model so that it can remain effective.

## III. Method

In this section, we have described the system setup and metrics for datasets. The assumptions for datasets are mentioned in appropriate locations. Similarly, the DevOps framework is described in detail. Each component of the framework is loosely coupled which results in a scalable system.

Since the dataset is healthcare-related, utmost care was taken to make sure all required guidelines set by NHANES are followed. There was no attempt to identify any personal information from the dataset.

### A. Dataset information

Data is collected from the National Health and Nutrition Examination Survey (NHANES). NHANES is conducted by the National Center for Health Statistics to assess the health and nutritional status. The survey data is from the civilian, non-institutionalized US population, and a multistage, stratified, clustered probability design was used. For this project, various data were collected which spanned 17 years (1999-2016).

**Independent Measures**

Demographics: Gender, age, race, country of birth, citizenship status, education level, marital status, income

Dietary: Energy intake in kcal, protein intake in grams, carbohydrate intake in grams, total sugars intake in grams, dietary fibers intake in grams, total fat intake in grams, and total saturated fatty acids intake in gram

Examination: systolic blood pressure, diastolic blood pressure, weight, body mass index, combined grip strength (kg), overall Oral Health Exam Status

Laboratory: Albumin (mg/L), Creatinine(umol/L), cholesterol, insulin, blood count 2.5 dimethylfuran(ng/mL).

Questionnaire: Average alcoholic drinks/day, worried/anxious status (how often), ever used drugs, health insurance status, emergency care visit for asthma, ever had a heart attack, taken prescription medicine, smoked at least 100 cigarettes in life, do you smoke now.

*Dependent Measures*

The number of times health care received in the past year is used as a proxy indicator for hospital utilization. This variable doesn't include times hospitalized overnight. If a patient is admitted to the hospital more than 5 times, then that patient is more susceptible to utilizing the hospital resources. With that assumption in mind, the patients are treated as highly probable to utilize hospital resources and low probability patients respectively.

B.  DevOps Framework Information

The first component of this framework is the code and build block. The code is hosted in a version control system(git). Data Scientists(DS) and ML engineers(MLE) can collaborate effectively via version control. For example- DS can enhance the analysis and learning capability of the end machine learning models, whereas the MLE can focus on the performance of the model by using various state-of-the-art model serving frameworks as mentioned in some recent studies [10], [11].

The build sub-component is responsible for packaging the code. "Code" refers to the development or programming phase, while "Build" represents the process of assembling or compiling that code into a deployable form (e.g., building containers or executables). A custom Github action framework is used for build which builds a docker container.

The second component consists of Test and deploy blocks. Once the build is ready, it's moved into a testing phase to verify functionality, correctness, and reliability. The testing phase is focused on mainly 2 parts. First is functional or unit testing to ensure the resulting code is executable and there is no logical issue. However, since we are dealing with a machine-learning model in this system, we need to ensure model correctness as well. The validation or hold-out dataset is essential in this case. Also since the framework supports Online Data as well, it's essential to utilize the test framework for continuous deployment as well. In this case, we have utilized the pyest (pytest.org) framework for testing. The model testing is written via custom Python code.

After tests pass, the new release is "Deployed" to the target environment so it can be used in the next stage. The deploy block is one of the most important parts of the whole framework. Once both code and models are ready, deploy logic kicks off and copies the required config and executable in the cloud providers for the next step. There are a few industry standard tools that were evaluated such as Jenkins, CircleCI and finally GitHub action was chosen given its simplicity.

However- since the framework is tool-agnostic, any components can be swapped with a compatible system. For example, the testing and build framework can be changed with javascript testing and TravisCI respectively without any loss in functionality.

Component 3 is the heart of the system which hosts the ML Model and also runs the Train-Validation-Test Cycle. The "NHANES data" (under "BatchData") feeds into the pipeline, providing a large dataset for building or training the ML model. The model then receives inputs (e.g., from the newly deployed code as well as from the NHANES data or other live sources) and outputs predictions or analyses. Complete architecture is provided in Figure 1.

The API serving system used in this component was FastAPI(fastapi.tiangolo.com). The model was built using the sci-kit-learn Python library. Also, the model was serialized and bundled with serving API. Online Data change can be done via a cron schedule trigger where a system periodically polls for new data and triggers the training pipeline. Alternatively, the framework can also support a push model where a data provider can send a notification to the component via an event system like Rabbit Message Queue or Kafka which will trigger the pipeline. Once again, the framework is not dependent on any tools or system, the information regarding tools is given here just as a reference architecture.

IV.  RESULTS AND ANALYSIS

Analysis is performed on the data from 1999-2000 to 2015-2016. Wald chi-square and t-tests were used to compare demographic, dietary, examination, laboratory, and questionnaire data with hospital utilization. Controlling for the above-mentioned variables, multiple logistic regression was used to create the models. Initially, models were constructed using a forward selection method and each class of covariates was added to evaluate the additional contribution from independent measures covariates. Models including all of the covariates were further analyzed using backward elimination. Missing values are dropped from the analysis in some cases (more than 5% missing) and imputated for other cases. Polynomial regression was used to impute categorical variables and predictive mean matching was used for numerical variables.

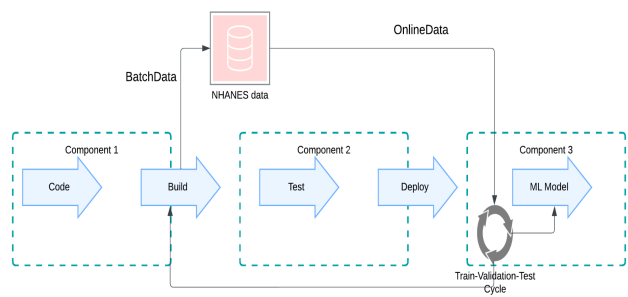

Fig. 1. DevOps Components and Workflow

Any further assumptions are notated in the respective result and analysis section.

*A. Factors predict the likelihood of a patient utilizing hospital resources.*

In traditional logistic regression models, we generally use odds ratio (OR) to assess the association between response and predictors. Usually, OR is considered an approximation of the Risk ratio (RR). However, since in this case, the outcome is not relatively rare (more than 10%), RR from the multivariate logistic regression [12] was calculated. In this analysis, only data from NHANES 2015-2016 is considered (n=9971). After running the stepwise model, the following variables are selected as the most significant variables. Standard errors and p-values are shown for the model in the below table. A P value <0.05 was considered statistically significant. Statistically significant variables are marked with *.

Age (0-18, 18-60, above 60*), Race(White, Non-Hispanic Black*, Other*), Citizenship(US, Non-US), Insurance(Covered, Not-covered*, others), Emergency care visit(Yes, No*, Missing), Heart attack(Yes, No, Others), Prescription drug use(Yes, No*, Others), Income, Total energy intake in kcal, Sugar intake, saturated fat intake, weight*, body-mass index(BMI)*, Albumin-urine, Insulin, Anxiety status(at least once in a month, a few times in year, never, others).

From the above result, it can be seen that demographic dietary, and health examination characteristics are key risk factors for a patient utilizing hospital resources. The details are shown in Table I.

TABLE I. HOSPITAL RESOURCE UTILIZATION PREVALENCE AND RELATIVE RISK BASED ON CHARACTERISTICS

|  | *Characteristics* | *Prevalence (%)* | *Relative risk (95% CI)* |
|---|---|---|---|
| Age group | 0-17 | 32.55 | 1.00 |
|  | 18-60 | 35.54 | 1.26(1.04-1.53) |
|  | 60+ | 31.91 | 1.56(1.28-1.90) |
| Race | White | 38.09 | 1.00 |
|  | Non-Hispanic Black | 18.09 | 0.66(0.57-0.77) |
|  | Other | 42.96 | 0.79(0.70-0.90) |
| Citizen | US | 93.12 | 1.39(1.12-1.75) |
|  | Non-US | 6.88 | 1.00 |
| Dietary characteristics | Energy Intake(Kcal) | N/A | 0.99(0.99-1.02) |
|  | Sugar Intake(gram) | N/A | 1.00(0.99-1.02) |
|  | Saturated Fat intake(gm) | N/A | 1.00(0.99-1.02) |
| Health characteristics | Weight | NA | 0.97(0.97-0.98) |
|  | Body-Mass Index | NA | 1.07(1.05-1.09) |
| Prescription Drug Use | Yes | 69.37 | 1.00 |
|  | No | 30.47 | 0.25(0.22-0.29) |
|  | Other | 0.16 | 1.24(0.24-5.24) |
| Anxiety Status | At least once a month | 48.99 | 1.00 |
|  | A few times in a year | 31.00 | 0.78(0.69-0.89) |
|  | Never | 19.58 | 0.77(0.67-0.89) |
|  | Other | 0.43 | 1.30(0.49-3.24) |
| Health Insurance | Yes | 93.97 | 1.00 |
|  | No | 5.87 | 0.52(0.42-0.65) |
|  | Other | 0.16 | 0.97(0.21-3.16) |

Hospital Resource utilization prevalence and Relative Risk based on Characteristics are given in Table I. As we can see, Age>60 years, US citizens, no prescription drug use, low anxiety status, and no insurance are the significant univariate risk predictors for hospital utilization. However, the hypothesis for risk factors due to Race, Citizenship, Prescription drug use, and Health insurance status needs to be analyzed carefully. As we are dealing with hospital utilization, access to hospital resources is somewhat limited for other races [13] which can explain the relatively low risk ratio for other races compared to white race. Citizenship and prescription drug use have a high correlation as far as hospital utilization is concerned. Similarly, limited access to health insurance can restrict access to hospital resources and thus the low-risk ratio due to "no health insurance" is expected.

With the above points in mind, we can conclude that age>60, body weight, BMI, dietary preferences such as high energy intake, more sugar and fat consumption, and high anxiety are the key risk factors that can predict the likelihood of a patient utilizing hospital resources.

*B. Different risk factors associated with different clinical categories*

For this part, survey responders are clustered into 3 different clinical categories- diabetes, obesity, and cardiovascular disease. Each of these categories was then analyzed using the similar process explained above. In this case, also only NHANES 2015-2016 is considered (n=9971).

**Diabetes**

In the case of diabetes clinical categories (P<0.05 marked as *) following predictors are used- Gender, Race (White, Non-Hispanic, black*, others), Marital Status (married, single, separated, others), Health insurance(Covered, Not Covered), Emergency care visit(Yes, No*, Missing), Heart attack(Yes, No, Others), Prescription drug use(Yes, No*, Others), Total carbohydrate intake, Albumin-urine, Anxiety status(at least once in a month, a few times in year, never, others). Compared to overall risk factors as mentioned in section a, we see age is not a significant factor for diabetes clinical categories. Also, marital status was not a significant risk factor for overall data, but it's a risk factor for this category. Apart from that, "Carbohydrate intake" dietary characteristic is a significant risk factor here which was not present in overall risk factors. Risk Ratio and prevalence details are given in Table II. From that table, one interesting observation is that a subject with gender as female and marital status as separated are relatively higher risk factors in case of hospital utilization for diabetes clinical categories controlling for all other variables.

TABLE II. Hospital Resource Utilization Prevalence and Relative Risk((Diabetes clinical categories)

|  | Characteristics | Prevalence (%) | Relative risk (95% CI) |
|---|---|---|---|
| Gender | Male | 48.44 | 1.00 |
|  | Female | 51.66 | 1.06(1.00-1.11) |
| Race | White | 33.12 | 1.00 |
|  | Non-Hispanic Black | 20.62 | 0.90(0.83-0.95) |
|  | Other | 46.25 | 0.92(0.86-0.97) |
| Marital Status | Married | 54.69 | 1.00 |
|  | Not Married | 9.69 | 1.02(0.94-1.10) |
|  | Separated | 35.31 | 1.08(1.02-1.14) |
|  | Others | 0.31 | 1.05(0.59-1.89) |
| Dietary characteristics | Carbohydrate Intake | N/A | 0.99(0.99-1.02) |
| Health characteristics | Emergency Care Visit(yes) | 17.50 | 1.00 |
|  | Emergency Care Visit(No) | 82.50 | 0.94(0.88-1.00) |
|  | Heart Attack(yes) | 14.37 | 1.00 |
|  | Heart Attack(No) | 85.00 | 0.94(0.88-1.01) |
|  | Other | 0.62 | 0.92(0.86-0.99) |
| Anxiety Status | At least once a month | 43.12 | 1.00 |
|  | A few times in a year | 34.06 | 0.97(0.92-1.03) |
|  | Never | 19.58 | 0.92(0.86-0.97) |
|  | Other | 0.43 | 1.23(0.89-1.63) |
| Health Insurance | Yes | 94.69 | 1.00 |
|  | No | 5.31 | 0.89(0.82-0.96) |
|  | Other | 0.01 | 0.74(0.30-1.49) |

**Obesity**

Obesity is defined as an individual who has a BMI>30. The relative risk factors details are given in Table III.
Compared to overall risk factors, the obesity risk factors, and relative risk follow diabetes clinical categories closely. Hence, there is a correlation between obesity and diabetes. Also in contrast with diabetes, smoking and sugar intake are significant factors in this case.

TABLE III: Hospital Resource utilization prevalence and Relative Risk (Obesity clinical categories)

|  | Characteristics | Prevalence (%) | Relative risk (95% CI) |
|---|---|---|---|
| Gender | Male | 64.8 | 1.00 |
|  | Female | 35.2 | 1.05(1.00-1.10) |
| Race | White | 38.82 | 1.00 |
|  | Non-Hispanic Black | 21.71 | 0.89(0.83-0.95) |
|  | Other | 39.47 | 0.92(0.86-0.97) |
| Citizen | US | 93.59 | 1.00 |
|  | Non-US | 6.41 | 1.03(0.98-1.07) |
| Dietary characteristics | Sugar Intake | N/A | 1.00(0.99-1.01) |
| Health characteristics | Heart Attack(yes) | 10.36 | 1.00 |
|  | Heart Attack(No) | 87.34 | 0.87(0.83-0.92) |
|  | Heart Attack(Other) | 2.30 | 0.96(0.76-1.20) |
| Prescription Drug Use | Yes | 87.83 | 1.00 |
|  | No | 12.01 | 0.81(0.74-0.92) |
| Anxiety Status | At least once a month | 50.99 | 1.00 |
|  | A few times in a year | 29.77 | 0.95(0.92-1.03) |
|  | Never | 19.08 | 0.96(0.86-0.97) |
|  | Other | 0.16 | 0.91(0.63-1.26) |
| Smoking (do you smoke now) | Yes | 9.70 | 1.00 |
|  | Never | 76.48 | 0.98(0.94-1.03) |
|  | Others | 13.82 | 0.95(0.91-1.00) |
| Health Insurance | Yes | 93.75 | 1.00 |
|  | No | 5.31 | 0.89(0.82-0.96) |
|  | Other | 0.01 | 0.74(0.30-1.49) |

**Cardiovascular health**

Apart from the common risk factors in the other 2 clinical categories, we see calorie intake, carbohydrate intake, and saturated fat intake are significant factors here. Also- age is an important factor here and the risk of cardiovascular health-related hospital utilization is significantly higher for the 60+ age group. Moreover, no prescription drug use has only 0.22 risk factors which is comparatively less than the other 2 clinical categories(i.e. diabetics and obesity). The risk factors details are notated in Table IV.

TABLE IV: Hospital Resource utilization prevalence and Relative Risk (CardioVascular health)

| | Characteristics | Prevalence (%) | Relative risk (95% CI) |
|---|---|---|---|
| Age group | 18-60 | 37.37 | 1.00 |
| | 60+ | 62.63 | 1.18(0.98-1.40) |
| Race | White | 45.96 | 1.00 |
| | Non-Hispanic Black | 18.18 | 0.79(0.62-0.99) |
| | Other | 35.86 | 0.94(0.86-0.97) |
| Dietary characteristics | Energy Intake(Kcal) | N/A | 1.00(0.99-1.01) |
| | Sugar Intake(gram) | N/A | 1.00(0.99-1.01) |
| | Saturated Fat intake(gm) | N/A | 1.00(0.99-1.01) |
| Health characteristics | Heart Attack(yes) | 23.74 | 1.00 |
| | Heart Attack(No) | 75.76 | 0.76(0.62-0.92) |
| | Heart Attack(Other) | 0.51 | 0.59(0.15-1.52) |
| Prescription Drug Use | Yes | 94.95 | 1.00 |
| | No | 5.05 | 0.22(0.74-0.92) |
| Anxiety Status | At least once a month | 61.36 | 1.00 |
| | A few times in a year | 26.52 | 0.95(0.92-1.03) |
| | Never | 11.36 | 0.96(0.86-0.97) |
| | Other | 0.76 | 0.91(0.63-1.26) |
| Smoking Status(lifetime cigarette) | More than 100 | 52.02 | 1.00 |
| | Less than 100 | 47.73 | 0.88(0.94-1.03) |
| | Others | 0.25 | 1.2(0.12-7.78) |
| Health Insurance | Yes | 93.94 | 1.00 |
| | No | 6.06 | 0.89(0.82-0.96) |

### C. Risk factors over time

To understand how the risk factors are changed over time, four sets of NHANES data were considered (1999-2000, 2005-2006, 2009-2010, 2015-2016). From this analysis, we observed that Age, Energy intake, health insurance status, ever had a heart attack, health insurance status, prescription drug use, BMI and weight are significant risk factors throughout the years. Gender, Race, sugar intake, insulin, and blood pressure are also common risk factors but these are not present in all years. Moreover, anxiety status has become a significant risk factors over the last 15 years, but it was not the case for 1999-2000 data.
Similarly, smoking is also a significant factor for both 2005-2006 and 2009-2010 data. Surprisingly overall oral health exam status is a common significant factor for both 1999-2000 and 2005-2006 data. We hypothesize that generally, patients who visit hospitals for oral health care are the ones who go to the hospital more often which increases the overall hospital resource usage. These changes in risk factors are described in Table V. The 'x' marks denoted if the risk factor was prevalent.

TABLE V: HOSPITAL RESOURCE UTILIZATION PREVALENCE - RISK FACTOR OVER TIME

| Characteristics | 1999-00 | 2005-06 | 2009-10 | 2015-16 |
|---|---|---|---|---|
| Age group | x | x | x | x |
| Race | | x | x | x |
| Gender | x | x | x | |
| Income | | | | x |
| Education | | | x | |
| Marital Status | | x | | |
| Energy Intake | x | x | x | x |
| Carbohydrate Intake | x | | | |
| Sugar Intake | | x | x | x |
| Fiber Intake | | | x | |
| Fat Intake | x | | x | x |
| Health Insurance | x | x | x | x |
| Ever Had a Heart Attack | x | x | x | x |
| Emergency Care Visit | | x | x | x |
| Prescription Drug Use | x | x | x | |
| Anxiety Status | | x | x | x |
| Smoking Status | | x | x | |
| Overall Oral Health Exam | x | | x | |
| Blood Pressure | x | x | x | |
| Weight | x | x | x | x |
| BMI | x | x | x | x |

### D. Utilization of hospital resources with increased prevalence of certain diseases

The prevalence of certain diseases (e.g. diabetes, obesity) has been thought to be increasing over time. As disease prevalence increases, has the use of hospital resources also increased? From the NHANES study [1] and other sources [14], it's being observed that the disease prevalence is increasing.
Compared to 2000, 2005, and 2010 both had significantly more hospital admissions in both 4-9 admissions per 12-month category and more than 10 admissions per 12-month category. For 2015, there is a slight decrease in 4-9 admissions per 12-month category. However, as shown in Figure 2, it has the highest number of patients in more than 10 admissions per 12-month category.

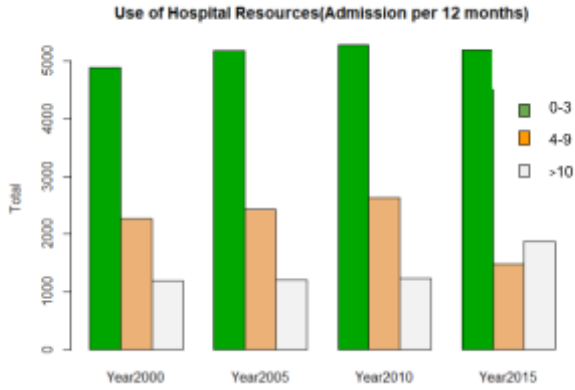

Fig. 2. Utilization of hospital resources *(Admission per 12 months)*

### E. Socio-economic factors associated with increases in disease prevalence

Socio-economic factors [15] include income, education, occupation, and insurance status. Apart from that we also used overall self-rating for health, and poverty income ratio (PIR) as measures for socio-economic factors. As we can see from Table VI below, there is a positive relation between higher education and high PIR to hospital resource usage. This can be explained as people with these 2 characteristics have higher access to hospital usage. In contrast with the full model, this model puts significantly more emphasis on health insurance status. Not covered by health insurance has a risk ratio of 0.46 (covered by health insurance is referent here with risk ratio as 1.)

TABLE VI: Hospital Resource utilization prevalence - socio-economic factor

| | *Characteristics* | *Prevalence (%)* | *Relative risk (95% CI)* |
|---|---|---|---|
| Education | Up to High School | 52.82 | 1.00 |
| | College | 25.21 | 1.05(0.84-1.32) |
| | Graduate or More | 21.86 | 1.36(1.11-1.66) |
| PIR | Less than 1.2 | 36.51 | 1.00 |
| | 1.2-2.0 | 23.85 | 0.89(0.71-1.10) |
| | 2.1-3.3 | 18.10 | 0.75(0.58-0.98) |
| | More than 3.3 | 21.55 | 1.18(0.87-1.64) |
| Health Insurance | Yes | 92.78 | 1.00 |
| | No | 7.11 | 0.47(0.21-0.96) |

### Conclusion

Risk factors for hospital resource usage are diverse. However, we do see some common patterns for respective characteristics. The trend for hospital resource usage is upward. Apart from standard demographic characteristics such as gender, age, and race- the most important other factors are lifestyle-related. High energy food intake, smoking, more carbohydrate, fat, and sugar intake, and high insulin presence can predict the risk factors for hospital resource usage. Socioeconomic status does play a role in this case as well. Keeping all these points in mind, our recommendation is to encourage the general population towards a healthier lifestyle to lower the risk of hospital resource usage.

However, the absence of a comprehensive system for analyzing this data in an accessible way hinders complex analysis. Moreover, the data is not static and the framework should be capable of updating itself without significant change. These challenges pose risks to utilizing these datasets to their full extent.

As the DevOps framework proposed here is capable of serving both batch and online data, this system can be useful to address all the shortages of current processes. Moreover, the framework proposed in this study is generic and thus can be used for various domains beyond healthcare. The components mentioned here are modular and suitable for any system where complex analysis is required. In the future, the framework can be utilized with synthetic data to enhance the capability of the system where the data is sparse such as in environmental science, robotics, cybersecurity cultural heritage, and arts.

### Acknowledgment

The authors would like to thank the CDC and NCHS for providing the NHANES data which is essential for this study.